\documentclass{article}

\usepackage{spconf,amsmath,graphicx}
\usepackage[table]{xcolor}
\usepackage{underscore}
\usepackage{xspace}
\usepackage{enumitem}

\usepackage{titlesec}
\titlespacing*{\section}{0pt}{1.2ex plus 1ex minus .2ex}{1.2ex plus .1ex}
\titlespacing*{\subsection}{0pt}{1.2ex plus 1ex minus .2ex}{1.2ex plus .2ex}

\ninept
\usepackage{hyperref}

\def\sigpos{\cellcolor{green!10}}
\def\signeg{\cellcolor{red!10}}
\def\open{\textbf{\underline{O}penness}\xspace}
\def\con{\textbf{\underline{C}onscientiousness}\xspace}
\def\ext{\textbf{\underline{E}xtraversion}\xspace}
\def\agr{\textbf{\underline{A}greeableness}\xspace}
\def\neu{\textbf{\underline{N}euroticism}\xspace}

\title{Comparing approaches for mitigating intergroup variability in personality recognition}

\name{Guozhen An$^{1,2}$, Rivka Levitan$^{1,3}$}
\address{$^1$Department of Computer Science, CUNY Graduate Center, USA \\
$^2$Department of Mathematics and Computer Science, York College (CUNY), USA \\
  $^3$Department of Computer and Information Science, Brooklyn College (CUNY), USA }

\begin{document}

\maketitle

\begin{abstract}
Personality have been found to predict many life outcomes, and there have been huge interests on automatic personality recognition from a speaker's utterance. Previously, we achieved accuracies between 37\%--44\% for three-way classification of high, medium or low for each of the Big Five personality traits (Openness to Experience, Conscientiousness, Extraversion, Agreeableness, Neuroticism). We show here that we can improve performance on this task by accounting for heterogeneity of gender and L1 in our data, which has English speech from female and male native speakers of Chinese and Standard American English (SAE). We experiment with personalizing models by L1 and gender and normalizing features by speaker, L1 group, and/or gender. 
\end{abstract}


\section{Introduction}

``Personality'' is defined by the \textit{Encyclopedia of Psychology}~\cite{kazdin2000encyclopedia} as ``individual differences in characteristic patterns of thinking, feeling and behaving,'' and is considered a primary source of inter-personal variation. The {NEO-FFI} five factor model of personality traits is the most commonly used model of personality, also known as the Big Five:  Openness to Experience (having wide interests, imaginative, insightful), Conscientiousness (organized, thorough, a planner), Extroversion (talkative, energetic, assertive), Agreeableness (sympathetic, kind, affectionate), and Neuroticism (tense, moody, anxious)~\cite{costa1992revised}.  These traits were originally identified by several researchers working independently~\cite{Digman1990} and the model has been employed to characterize personality in multiple cultures~\cite{McCrae2001}. 

Of particular interest to researchers is the potential for personality traits to account for or mitigate the interpersonal variation that makes deception detection a uniquely difficult task. Previous work has found personality-based differences in deception as well as deception detection~\cite{enos2006personality,Levitan2015cognitive}. This work, which is based on data collected in the context of a deceptive task, can be used to improve deception detection by incorporating automatically-detected information about personality.

Our corpus includes English speech from female and male speakers who are native speakers of Standard American English (SAE) and Mandarin Chinese (MC). We hypothesize that intra-group differences inhibit the performance of models trained on this heterogeneous data. Female and male speakers are known to have different vocal characteristics and use language differently (e.g.~\cite{levitan2016identifying,argamon2005lexical,shafran2003voice,zeng2006robust}). The same is true for speakers with different native languages (L1)~\cite{tetreault2013report}. Furthermore, additional intra-group variability may stem from the fact that personality can be expressed differently in different cultures~\cite{scherer1979personality}.

In our current work we experiment with ways to automatically identifying the NEO-FFI Big Five personality traits from speech, which will be useful for applications such as dialogue systems. Although there is previous research on this task, most has focused on predicting personality scores labeled by annotators asked to identify personality traits of others, rather than from self-reported personality inventories. Although ratings by observers who know the subject well are considered valid in personality research, ratings by strangers have been shown to correlate only weakly with self reports, and have moderate to weak internal consistency (as measured by Cronbach's alpha)~\cite{borkenau1992trait}.

We experiment with two different methods for accounting for this variation: homogenization (partitioning the training data into homogeneous models) and normalization. Homogenization ensures that we train and test on data of the same kind, with the disadvantage of less training data per model. Normalization accounts for differential vocal characteristics and uses of language, while allowing for the use of the entire dataset, but does not handle differential expressions of personality. We compare the two approaches for each personality trait and gender/L1 group and show that partitioning the data into homogeneous models improves performance in most cases. 

\section{Related Work}

Numerous studies have successfully predicted \emph{observer} reports of personality from speech and text \cite{an2015literature}. Mohammadi et al.\cite{mohammadi2010voice} used prosodic features to detect personality from ten second audio clips labeled for personality by human judges based on the audio clips alone, and reported recognition rates ranging from 64.7\% (\agr) to 79.4\% (\ext) for the binary classification problem derived from splitting personality scores at the mean.

Argamon et al.~\cite{argamon2005lexical} used stylistic lexical features to classify student essays as high or low (top or bottom third) for \ext and \neu. 
Linguistic Inquiry and Word Count (LIWC)~\cite{pennebaker2001linguistic} categories, have been shown to correlate with Big Five personality traits, both in writing samples~\cite{pennebaker1999linguistic} and in spoken dialogue~\cite{mehl2006personality}.

In a very comprehensive study, Mairesse et al. predicted self and observer personality scores from essay and conversational data, using LIWC, psycholinguistic, and prosodic feature sets. Their results indicate that observer reports are easier to predict, they achieved good results with models of observed personality but no results above baseline with models of self-reported personality.

In previous work~\cite{an2016automatically}, we predicted \emph{self}-reports of personality scores using acoustic-prosodic and language features and found that the best performance for each trait was achieved by models with different combinations of feature sets. Performance ranged from 37-44\%, significantly above the baseline for each trait. 

This work similarly predicts \emph{self}-reports, a more difficult task than predicting stranger/observer reports. Although ratings by observers who know the subject well are considered valid in personality research, ratings by strangers have been shown to correlate only weakly with self reports, and have moderate to weak internal consistency (as measured by Cronbach's alpha)~\cite{borkenau1992trait}. Additionally, the HI/ME/LO classes we predict are determined by population norms, a more motivated criterion than the naive statistical thresholds used by most studies. 

Our hypothesis that intra-group differences in gender and L1 affect personality expression is motivated by Scherer's survey of personality markers in speech~\cite{scherer1979personality}, which cites studies showing that f0 (pitch), for example, is associated with dominance in American males, discipline in German males, and dullness in American females, among other similar findings. 

\section{Data}

The collection and design of the corpus analyzed here is described in more detail by~\cite{Levitan2015cognitive,Levitan2015wmdd,levitan2016identifying}. It contains within-subject deceptive and non-deceptive English speech from native speakers of Standard American English (SAE) and Mandarin Chinese (MC), with native language defined as the language spoken at home until age five. There are approximately 125 hours of speech in the corpus from 173 subject pairs and 346 individual speakers. 

The data was collected using a fake resume paradigm, where pairs of took turns interviewing their partner and being interviewed from a set of 24 biographical questions such as ``What is your mother's job?'' Subjects were instructed to lie in their answers to a predetermined subset of the questions. The interviews took place in in a soundproof booth and the subjects were recorded by close-talk headsets. 

Before the recorded interviews, subjects filled out the NEO-FFI (Five Factor) personality inventory~\cite{costa1992revised}, yielding scores for Openness to Experience (O), Conscientiousness (C), Extraversion (E), Agreeableness (A), and Neuroticism (N).

We also collected a 3-4 minute baseline sample of speech from each subject for use in speaker normalization, in which the experimenter asked the subject open-ended questions (e.g., What do you like the best/worst about living in NYC?). Subjects were instructed to be truthful in answering. Once both subjects had completed all the questionnaires and we had collected both baselines, they began the lying game. 

Transcripts for the recordings were obtained using Amazon Mechanical Turk\footnote{\url{https://www.mturk.com}} (AMT). Three transcripts for each audio segment from different `Turkers' were obtained, and combined using {\it rover} techniques~\cite{fiscus1997post}, producing a rover output score measuring the agreement between the initial three transcripts. For clips with a score lower than 70\% (9.7\% of the clips), transcripts were manually corrected.

\section{Method}

\subsection{Features}

We use the feature sets described in our previous work~\cite{an2016automatically}, which include acoustic-prosodic low-level descriptor features (\textbf{LLD}); word category features from \textbf{LIWC} (Linguistic Inquiry and Word Count)~\cite{pennebaker2001linguistic}; and word scores for pleasantness, activation and imagery from the Dictionary of Affect in Language (\textbf{DAL})~\cite{whissell1986dictionary}. We also add two new feature sets, word vectors (\textbf{WV}) and part-of-speech counts (\textbf{POS}).

\textbf{Word vectors (WV)}. Continuous vector representations of words (word vectors) have been used in statistical language modeling~\cite{bengio2003neural}, speech recognition~\cite{schwenk2007continuous} and a wide range of NLP tasks~\cite{glorot2011domain} with considerable success. Motivated by these findings, we use the Gensim library~\cite{rehureklrec} to extract word vector features using Google's pre-trained word vector model~\cite{mikolov2013distributed}. In order to calculate the vector representation of turn level segment, we extract a 300 dimensional word vector for each word of the turn segment, and then average them to get a 300 dimensional vector representation of the entire turn segment. 

\textbf{Part-of-speech counts (POS).} Previous research has shown that Part-Of-Speech (POS) features predict personality effectively~\cite{wright2014personality}. In order to capture this information, we extract 45 POS features using Natural Language Toolkit \cite{bird2006nltk} with the Stanford POS tagger~\cite{toutanova2003feature}. We count the occurrence of 45 POS types from each turn in the corpus.

\subsection{Homogeneous models}

\begin{figure}[t]
   \centering
   \includegraphics[width=\linewidth]{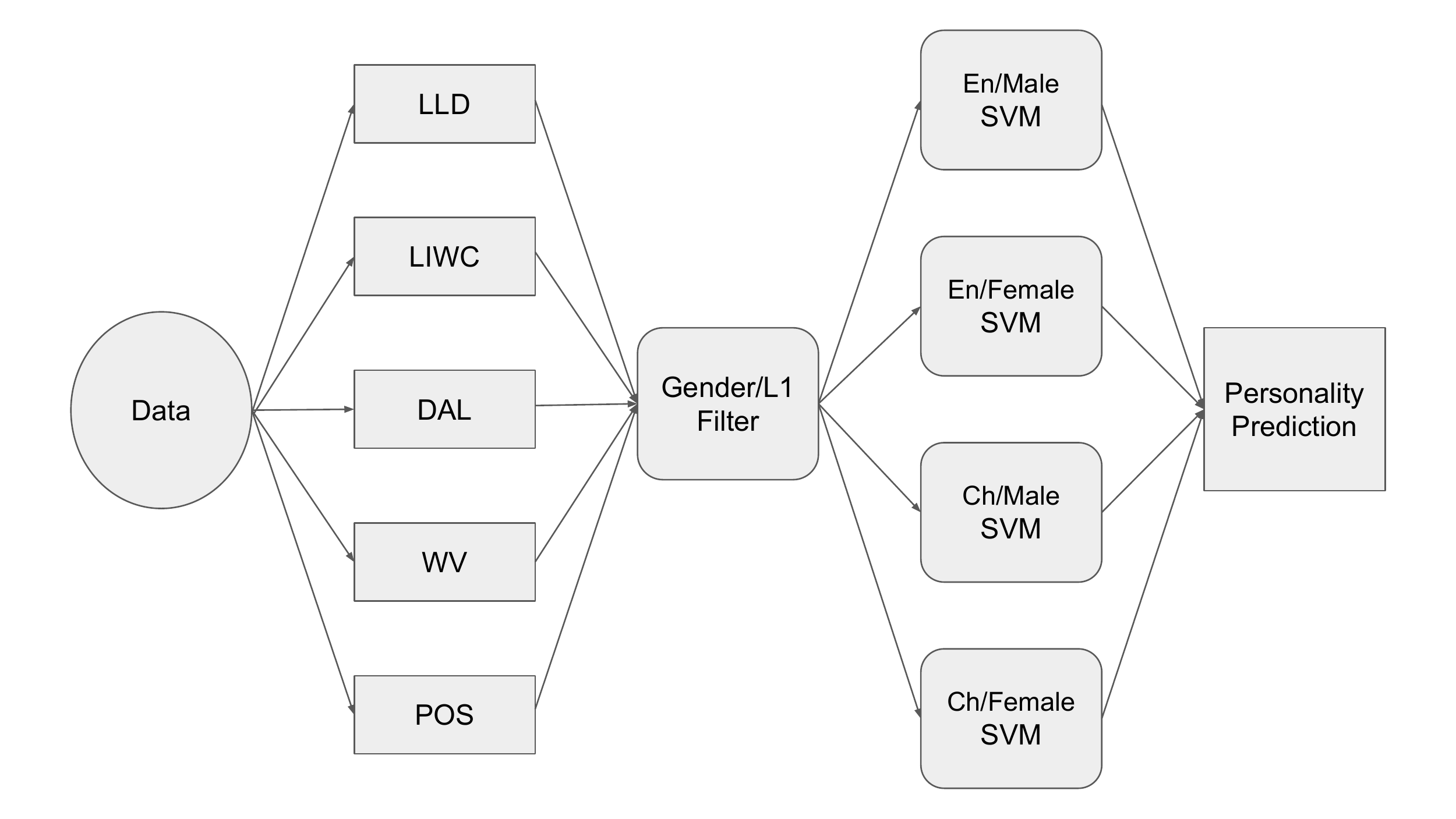}
   \caption{{\it Homogeneous model.}}
   \label{fig:Homogeneous}
\end{figure}

Our data includes approximately equal amounts of English speech from female and male speakers of Standard American English (SAE) and Mandarin Chinese (MC). In order to mitigate the effect of gender- and L1-based acoustic-prosodic and language variation, as well as variation in the expression of personality, we train four separate models for each personality trait, each using half the data and homogeneous with respect to either gender or L1 (Figure~\ref{fig:Homogeneous}). That is, we train a model using only female speech, another using only male speech, and two more using only speech from native SAE or MC speakers, respectively. 

At test time, the personality of each speaker is predicted using the corresponding homogeneous model. We experiment with with both single-gender and single-L1 models. To evaluate the effectiveness of this technique for mitigating data heterogeneity, we match speakers to models using the gold-standard gender and L1 labels recorded during data collection. In an online application where such labels might not be available, they could be automatically predicted; accuracy for gender prediction is as high as 95\% in deployed systems~\cite{levitan2016automatic,shafran2003voice}, while L1 detection has accuracies of about 80\% on essay data~\cite{brooke2013native,tetreault2013report}.

\subsection{Normalization}

\begin{figure}[t]
   \centering
   \includegraphics[width=\linewidth]{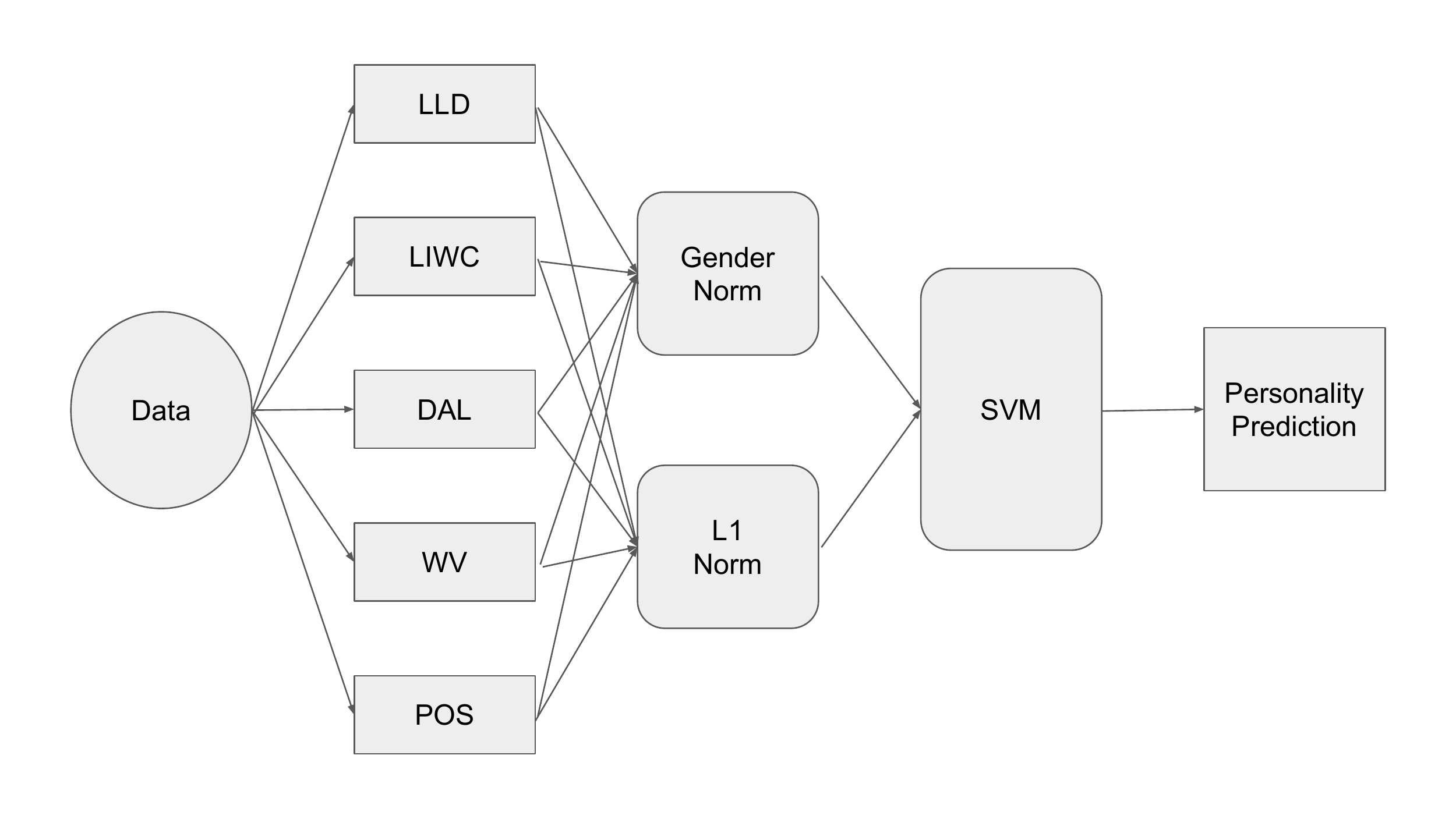}
   \caption{{\it Normalization model.}}
   \label{fig:Normalization}
\end{figure}

The disadvantage of partitioning the training data into homogeneous models is the consequent reduction in training data size (Figure~\ref{fig:Normalization}). We experiment with an alternate approach, using all the training data in a single model but first normalizing the data by speaker, gender, or L1. Normalization can account for population-characteristic differences in vocal qualities and language use, but does not directly mitigate differential expression of personality traits.

We use $z$-score normalization, which represents each value in terms of standard deviations from the population mean, so that a high-pitched female voice, for example, would be comparable to a high-pitched male voice, disregarding that the raw female pitch value may be as much as 200 Hz higher than the raw male pitch value. For each value $x$, the normalized value is \(\frac{x-\mu}{\sigma}\), where $\mu$ and $\sigma$ are the corresponding population mean and standard deviation, respectively.

We experiment with speaker, gender, and L1 normalization. For gender and L1 normalization, $\mu$ and $\sigma$ are calculated from the portion of the training data corresponding to that gender or L1, and used to normalize both the training and test data. For speaker normalization, $\mu$ and $\sigma$ are calculated separately for each speaker in both the training and test data. This corresponds to a likely use of these models, in which the target speaker's data is available but an entire population's is not.

\subsection{Machine learning experiments}

\cite{mairesse2007using} experimented with personality recognition as a classification, regression, and ranking problem, finding that ranking models performed best overall. We believe that classification is the best format for using personality scores in downstream applications such as deception detection or adapting dialogue systems. Rather than split scores into equal bins, we label the NEO scores as ``High''  (HI), ``Medium'' (ME) or ``Low'' (LO) for each dimension, based on thresholds derived from population norms from a large sample of administered NEO-FFI tests. 

We used Weka's SVM classifier for these experiments. The unit of classification used is the turn, which were extracted in the following manner: the manual orthographic transcription was force-aligned with the audio, and the speech was segmented if there was a silence of more than 0.5 seconds.

In total, there are approximately 30000 turn-level instances, which are upsampled during training so that the HI, ME and LO labels, which are normally distributed, will be balanced in the training set. The average duration of each instance is 3.77s, though there are some particularly long or short outliers.

To focus the analysis on the homogenization and normalization techniques discussed here, we combine all feature sets together, even though we previously found \cite{an2016automatically} that performance was higher for some traits using a subset of the features. Future work will explore more effective methods for combining feature sets. As in previous work, we predict each trait separately.

\section{Results}

\subsection{Baseline}
The baseline for these experiments is a model for each personality factor including all feature sets (LLD, LIWC, DAL, WV, POS) and all genders and L1s, without normalization. The accuracies for each trait are shown in Table~\ref{tab:baseline}. In this and all subsequent tables, highlighted results are significant at a two-tailed .95 confidence interval. The baseline surpasses a chance baseline of 0.33 (for a three-way classification of HI vs. MED vs. LO) for only \textbf{\underline{O}penness} and \textbf{\underline{N}euroticism}. The baseline \textbf{\underline{E}xtraversion} model performs \textit{worse} than chance. 

\begin{table}[h]
\caption{\label{tab:baseline} Baseline accuracy for each trait. Green and bold cells are significantly \textit{better} than chance (with 95\% confidence); red cells are significantly \textit{worse} than chance.}
\vspace{1ex}
\centering
\begin{tabular}{|l|l|l|l|l|l|}\hline
 & \textbf{O} & \textbf{C} & \textbf{E} & \textbf{A} & \textbf{N} \\ \hline
Chance & 0.33 &	0.33 & 0.33 & 0.33 & 0.33 \\ \hline
All Feat & \sigpos \textbf{0.39}&	0.33&	\signeg 0.31&	0.33&	\sigpos \textbf{0.35} \\ \hline
\end{tabular}
\end{table}

\subsection{Homogeneous models}
Table~\ref{tab:hom} shows the performances of the homogeneous models. The disparate results reinforce our previous observation~\cite{an2016automatically} that each personality factor can be considered a separate task requiring its own approach. 

\begin{table}[h]
\caption{\label{tab:hom} Accuracies for homogeneous models. Green and bold cells are significantly \textit{better} than the baseline model (with 95\% confidence); red cells are significantly \textit{worse} than the baseline model.}
\vspace{1ex}
\centering
\begin{tabular}{|l|l|l|l|l|l|}\hline
& \textbf{O} & \textbf{C} & \textbf{E} & \textbf{A} & \textbf{N} \\ \hline
male	&\signeg 0.32	&\signeg0.31	&0.31	& 0.34	&\signeg0.29\\
female	&\signeg0.34	&\signeg0.26	&\sigpos \textbf{0.35}	&0.32	&\sigpos \textbf{0.43}\\
chinese	&\signeg0.34	&\sigpos \textbf{0.37}	&\sigpos \textbf{0.35}	&\sigpos \textbf{0.39}	&\signeg0.26\\
english	&\signeg0.36	& 0.34	&\sigpos \textbf{0.35}	&\sigpos \textbf{0.36}	&0.34\\ \hline

male/chinese & \signeg0.35		&\sigpos \textbf{0.39}	&\signeg0.26	& 0.34	&\signeg0.25 \\
male/english & 0.38		&\sigpos \textbf{0.36}	&\sigpos \textbf{0.37}	&0.33	&\signeg0.32\\
female/chinese & \signeg0.35	&\sigpos \textbf{0.37}	&\sigpos\textbf{0.33}	&0.32	& 0.36\\
female/english & \sigpos \textbf{0.48}	&\sigpos \textbf{0.38}	&\sigpos \textbf{0.47}	&\signeg0.27	&\sigpos \textbf{0.40}\\\hline
\end{tabular}
\end{table}

Seven of the eight homogeneous models for \open performed worse than the baseline heterogeneous \open model, which had the highest accuracy (0.39) for this trait, as in our previous work~\cite{an2016automatically}. \neu, which had the second highest baseline (heterogeneous) accuracy, had worse performance for the male, Chinese L1, and male/L1 models, but a performance improvement of 8 percentage points for the female model, yielding an accuracy of 0.43, and 5 percentage points for the female/English model. \con had worse performance for the gender-homogeneous models, but improved for the L1-homogeneous models and gender/L1-homogeneous models. The female, Chinese, English and all the gender/L1-homogeneous \ext models except male/Chinese outperformed chance, which the corresponding baseline model failed to clear, as did the L1-homogeneous \agr models.

It is especially noteworthy that the Chinese-L1 models outperformed the heterogeneous baseline for \con, \ext and \agr. The subset of the corpus that was available at the time of analysis was not balanced with respect to L1, and the size of the Chinese-L1 training data was only 30-39\% of the heterogeneous training data (the exact percentage was different for each trait). That the Chinese-L1 models achieved better performance with approximately a \textit{third} of the training data indicates that the English-L1 instances did not contribute information useful for the classification of the Chinese-L1 instances, and that future work should not combine data from different L1 speakers indiscriminately. We might also suggest, based on these results, that the expression of \con, \ext and \agr is affected by cultural differences between native speakers of MC and SAE.
Conversely, the success of the heterogeneous \open model compared to the four corresponding homogeneous models suggests the interpretation that Openness is expressed similarly across cultural backgrounds. 

The results for the gender-homogeneous models are mixed. The male models do not outperform the baseline for any trait, and perform worse in most cases; the female models fall short of the baseline for \open and \con but outperform it significantly for \ext and \neu. It is unclear whether data from the other gender contributes valuable information in cases where the baseline dominates, or whether the benefits of a homogeneous model were not enough to outweigh the reduction in training data, which is 40-49\% for the male models.

When the data is broken down further to train models homogeneous with respect to both gender and L1, performance improves notably for \con for all four groups, and for all traits except \agr for English-speaking females. The improved performance in \open for female/English (10 percentage points) is the only improvement we achieve for that trait over all experiments. 

\subsection{Normalization}
Table~\ref{tab:norm} shows the results for speaker, gender and L1 normalization. 

\begin{table}[h]
\caption{\label{tab:norm} Accuracies for normalized models. Green and bold cells are significantly \textit{better} than the baseline model (with 95\% confidence); red cells are significantly \textit{worse} than the baseline model.}
\vspace{1ex}
\centering
\begin{tabular}{|l|l|l|l|l|l|}\hline
& \textbf{O} & \textbf{C} & \textbf{E} & \textbf{A} & \textbf{N} \\ \hline
Speaker & \signeg0.33	&0.33			&\sigpos\textbf{0.33}	&0.33	&\signeg0.33\\
Gender & 0.38		&\sigpos \textbf{0.35}	&	0.33	&0.33	&0.33\\
L1 & \signeg0.34	&\sigpos \textbf{0.36}	&	0.32	&0.32	&0.33\\ \hline
\end{tabular}
\end{table}

The \textbf{speaker}-normalized models do not outperform chance for any trait, though the \ext model does outperform the below-chance heterogeneous baseline; the baseline surpasses the \open and \agr models.
Of the \textbf{gender}-normalized models, only the \con model outperforms the baseline. The \open model has higher accuracy, but does not outperform its relatively strong baseline.
Similarly, the \textbf{L1}-normalized \con model outperforms its baseline; the \open model does worse than its baseline, and normalization does not affect the performance of the \ext, \agr and \neu models.

\subsection{Homogenization vs. normalization}

Table~\ref{tab:homvnorm} shows the differences in performance between homogeneous vs. normalized models. Since there are eight separate homogeneous models, each including a subset of the training data, we compare each normalization method with the \emph{best} homogeneous model and the \emph{average} performance on that trait. 

\begin{table}[h]
\caption{\label{tab:homvnorm} Performance differences between homogenized vs. normalized models.}
\vspace{1ex}
\centering
\begin{tabular}{|l|l|l|l|l|l|}\hline
& \textbf{O} & \textbf{C} & \textbf{E} & \textbf{A} & \textbf{N} \\ \hline
& \multicolumn{5}{c|}{Best homogeneous performance}\\\hline
Speaker & \sigpos\textbf{+0.15}	&\sigpos\textbf{+0.06}	&\sigpos\textbf{+0.14}	&\sigpos\textbf{+0.06}	&\sigpos\textbf{+0.11} \\
Gender &\sigpos \textbf{+0.10}	&\sigpos\textbf{+0.03}	&\sigpos\textbf{+0.14}	&\sigpos\textbf{+0.06}	&\sigpos\textbf{+0.10}\\
Language &\sigpos \textbf{+0.14}	&\sigpos\textbf{+0.03}	&\sigpos\textbf{+0.15}	&\sigpos\textbf{+0.06}	&\sigpos\textbf{+0.10}\\\hline
& \multicolumn{5}{c|}{Average homogeneous performance}\\ \hline				
Speaker &\sigpos \textbf{+0.04}	&\sigpos \textbf{+0.02}	&\sigpos \textbf{+0.02}	&0.00	&0.00\\
Gender & -0.01	& -0.01	&\sigpos \textbf{+0.02}	&0.00	&0.00\\
Language &\sigpos \textbf{+0.02}	& -0.01	&\sigpos \textbf{+0.03}	&\sigpos \textbf{+0.01}	&0.00\\\hline
\end{tabular}
\end{table}

For every normalization model, there is at least one homogeneous model, representing one subset of the population in the corpus, that significantly outperforms it. Furthermore, none of the normalization models significantly outperform the \emph{average} homogeneous performance, and several do significantly worse. As a technique, our analysis shows that breaking down the data into homogeneous models is preferable to normalization almost across the board, despite the diminution of the training data. An interpretation of this finding is that intra-group differences in the expression of personality are more detrimental to classification accuracy than differences in vocal characteristics or language use.

\section{Discussion and Conclusions}
This paper discusses two approaches for dealing with data heterogeneity, along two dimensions of heterogeneity, for five orthogonal personality traits. From our comparisons between each approach, we can conclude the following:
\begin{itemize}[noitemsep,nolistsep]
\item performance on this task can be improved by mitigating the heterogeneity of the training data
\item partitioning the data into homogeneous models works better for this purpose than normalization
\item partitioning with respect to both gender and L1 works better than partitioning along just one dimension.
\end{itemize}

More specific trends that we observe, which we present as hypotheses to be verified on other datasets rather than concrete conclusions, include:
\begin{itemize}[noitemsep,nolistsep]
\item \open is better predicted with a full heterogeneous dataset than with reduced homogeneous models.
\item Conversely, \con is best predicted with homogeneous models.
\item The personality of female native speakers of SAE is predicted better with homogeneous models.
\item The personality of native MC speakers is predicted better with homogeneous models.
\end{itemize}

In future work we will explore the fusion of these findings with more sophisticated machine learning models, including late-fusion techniques for combining the various feature sets and multi-task learning for modeling the prediction of the five personality traits together, and evaluate whether the gains achieved here are orthogonal to other methods for improving performance. 

\section{Acknowledgements}

We thank Julia Hirschberg, Andrew Rosenberg, Sarah Ita Levitan, and Michelle Levine. This work was partially funded by AFOSR FA9550-11-1-0120.

\bibliographystyle{IEEEbib}
\bibliography{mybib}

\end{document}